\documentclass[%
reprint,
superscriptaddress,
showpacs,
amsmath,amssymb,
prc,
floatfix, ]%
{revtex4-1}

\usepackage{color}
\usepackage{CJK}

\usepackage{graphicx}
\usepackage{dcolumn}
\usepackage{bm}
\usepackage[dvipdfmx,bookmarks=true,colorlinks,%
            citecolor=blue,linkcolor=blue,anchorcolor=blue,filecolor=blue,urlcolor=blue,%
           ]{hyperref}

\begin{document}


\title{Pseudospin symmetry in single particle resonances in spherical square wells}

\author{Bing-Nan Lu}
 \affiliation{State Key Laboratory of Theoretical Physics,
              Institute of Theoretical Physics, Chinese Academy of Sciences,
              Beijing 100190, China}
\author{En-Guang Zhao}
 \affiliation{State Key Laboratory of Theoretical Physics,
              Institute of Theoretical Physics, Chinese Academy of Sciences,
              Beijing 100190, China}
 \affiliation{Center of Theoretical Nuclear Physics, National Laboratory
              of Heavy Ion Accelerator, Lanzhou 730000, China}
\author{Shan-Gui Zhou}
 \email{sgzhou@itp.ac.cn}
 \affiliation{State Key Laboratory of Theoretical Physics,
              Institute of Theoretical Physics, Chinese Academy of Sciences,
              Beijing 100190, China}
 \affiliation{Center of Theoretical Nuclear Physics, National Laboratory
              of Heavy Ion Accelerator, Lanzhou 730000, China}

\date{\today}

\begin{abstract}
\begin{description}
\item[Background]  
The pseudospin symmetry (PSS) has been studied extensively for bound states.
Recently we justified rigorously that the PSS
in single particle resonant states is exactly conserved when the attractive
scalar and repulsive vector potentials of the Dirac Hamiltonian have the
same magnitude but opposite sign [\href{\doibase 10.1103/PhysRevLett.109.072501}
      {\bibfield  {journal} {\bibinfo {journal} {Phys. Rev. Lett.}}
       \textbf {\bibinfo {volume} {109}},\ \bibinfo {pages} {072501} 
       (\bibinfo {year} {2012})}].
\item[Purpose]     
To understand more deeply the PSS in single particle resonant states,
we focus on several issues related to the exact 
conservation and breaking mechanism of the PSS in single particle resonances.
In particular, we are interested in how the energy and width splittings
of PS partners depend on the depth of the scalar and vector potentials.
\item[Methods]     
We investigate the asymptotic behaviors of radial Dirac wave functions.
Spherical square well potentials are employed
in which the PSS breaking part in the Jost function can be well isolated.
By examining the zeros of Jost functions corresponding 
to small components of the radial Dirac wave functions,
general properties of the PSS are analyzed.
\item[Results]     
By examining the Jost function, the occurrence of intruder orbitals is 
explained and it is possible to trace continuously the PSS partners 
from the PSS limit to the case with a finite potential depth.
The dependence of the PSS in resonances as well as in bound states on the
potential depth is investigated systematically.
We find a threshold effect in the energy splitting and 
an anomaly in the width splitting of pseudospin partners 
when the depth of the single particle potential varies from zero to a finite value.
\item[Conclusions] 
The conservation and the breaking of the PSS in resonant states and 
bound states share some similar properties. 
The appearance of intruder states can be explained by examining the zeros of Jost functions.
Origins of the threshold effect in the energy splitting 
and the anomaly in the width splitting of PS partners,
together with many other problems, 
are still open and should be further investigated.
\end{description}
\end{abstract}

\maketitle


\section{Introduction}

More than 40 years ago, the pseudospin symmetry (PSS) was found to be 
approximately conserved in atomic nuclei and 
it was shown that doublets of single particle levels with quantum numbers
$(n_{r},l,j=l+1/2)$ and $(n_{r}-1,l+2,j=l+3/2)$ in the same major shell 
are nearly degenerate~\cite{Arima1969_PLB30-517,Hecht1969_NPA137-129}.
Based on the pseudospin concept, a simple but useful pseudo-SU(3) model was 
proposed and
later this model was generalized to be the pseudo-symplectic 
model~\cite{Ratna-Raju1973_NPA202-433, Troltenier1994_NPA576-351, *Troltenier1995_NPA586-53}.
Since the PSS was observed, several nuclear phenomena have been 
interpreted in connection with the PSS, 
such as nuclear superdeformed configurations~\cite{Dudek1987_PRL59-1405, Bahri1992_PRL68-2133}, 
identical bands~\cite{Nazarewicz1990_NPA512-61, Zeng1991_PRC44-R1745},
and pseudospin partner bands~\cite{Xu2008_PRC78-064301, Hua2009_PRC80-034303}.
The PSS may also manifest itself in magnetic moments and 
transitions~\cite{Troltenier1994_NPA567-591, 
Ginocchio1999_PRC59-2487, *Neumann-Cosel2000_PRC62-014308}
and $\gamma$-vibrational states in atomic nuclei~\cite{Jolos2012_PRC86-044320}.
It is thus an interesting topic to investigate the origin and breaking 
mechanism of the PSS which has been done within various backgrounds.
With these studies a much better understanding of the nuclear structure 
based on the PSS is anticipated.

In early years much efforts were devoted 
to revealing connections between the normal spin-orbit representation 
and the ``pseudo'' spin-orbit one and 
to exploring the microscopic origin of the PSS 
with spherical, axially deformed, and triaxially deformed potentials
of (mostly) the oscillator type~\cite{Bohr1982_PS26-267, Castanos1992_PLB277-238, 
Bahri1992_PRL68-2133, 
Blokhin1995_PRL74-4149, *Blokhin1996_JPA29-2039, *Blokhin1997_NPA612-163, *Beuschel1997_NPA619-119}.
It was found that the PSS conserves almost exactly for an oscillator potential 
with one-body orbit-orbit ($v_{ll}$) and spin-orbit ($v_{ls}$) interaction 
strengths satisfying the condition $v_{ls} \approx 4 v_{ll}$;
moreover, this condition is consistent with relativistic mean-field 
results~\cite{Bahri1992_PRL68-2133}.
A big step towards the understanding of the origin of the PSS in atomic nuclei
was made in 1997 when Ginocchio revealed that 
the PSS is essentially a relativistic symmetry of 
the Dirac Hamiltonian and 
the pseudo orbital angular momentum $\tilde{l}$ is nothing but
the angular momentum of the small component of a Dirac spinor~\cite{Ginocchio1997_PRL78-436}.
It was shown that the PSS in nuclei is exactly conserved when the scalar potential $S(r)$
and the vector potential $V(r)$ have the same size but opposite sign, 
i.e., $\Sigma(r)\equiv S(r)+V(r)=0$~\cite{Ginocchio1997_PRL78-436}.
Later Meng et al. found that the PSS is connected with the competition 
between the centrifugal barrier and the pseudospin-orbit potential 
and the PSS is exact under the condition $d\Sigma(r)/dr=0$~\cite{Meng1998_PRC58-R628}.
This condition means that the PSS becomes much better for exotic nuclei 
with a highly diffused potential~\cite{Meng1999_PRC59-154}.
However, in either limit, $\Sigma(r)=0$ or $d\Sigma(r)/dr=0$, there are no bound states 
any more, thus in realistic nuclei the PSS is always broken.
In this sense the PSS is usually viewed as a dynamical symmetry~\cite{Arima1999_RIKEN-AF-NP-276,
Alberto2001_PRL86-5015}.
Following discussions for spherical nuclei, the study of the PSS within
the relativistic framework was 
quickly extended to deformed ones~\cite{Lalazissis1998_PRC58-R45, 
Sugawara-Tanabe1998_PRC58-R3065,*Sugawara-Tanabe1999_PRC60-019901}.
One consequence of the fact that the PSS is a relativistic symmetry of 
the Dirac Hamiltonian is that the relativistic wave functions of 
the corresponding pseudospin doublets satisfy certain relations which have 
been tested both in spherical and in deformed nuclei~\cite{Ginocchio1998_PRC57-1167, 
*Ginocchio2002_PRC66-064312,
Lalazissis1998_PRC58-R45, Sugawara-Tanabe2002_PRC65-054313, Ginocchio2004_PRC69-034303}.

By solving the Dirac Hamiltonian, one gets not only positive energy
states in the Fermi Sea, but also negative energy states in the Dirac sea. 
When solutions of the Dirac Hamiltonian are used as a complete basis,
e.g., in the Dirac Woods-Saxon basis,
states with both positive and negative energies must be 
included~\cite{Zhou2003_PRC68-034323,*Zhou2006_AIPCP865-90,*Zhou2008_ISPUN2007,
Zhou2010_PRC82-011301R,*Li2012_PRC85-024312,*Li2012_CPL29-042101,*Li2012_AIPCP1491-208,
*Li2013_AIPCP1529-190,
Chen2012_PRC85-067301, Schunck2008_PRC77-011301R,*Schunck2008_PRC78-064305,
Long2010_PRC81-024308}.
Negative energy states correspond to anti-particle states.
In both cases, there are discrete bound states and continuum states.
The PSS study has been generalized not only from the Fermi sea to the Dirac sea, 
i.e., from single particle states to anti-particle states, but also 
from bound states to continuum states.
The PSS in negative energy states means the spin symmetry (SS) in anti-nucleon 
spectra~\cite{Ginocchio1999_PR315-231, Zhou2003_PRL91-262501}.
The SS in single anti nucleon energy spectra was explored
and found to be much better developed than the PSS in 
normal nuclear single particle spectra~\cite{Zhou2003_PRL91-262501}.
The SS in antinucleon spectra was also tested by investigating relations
between Dirac wave functions of spin doublets with the relativistic mean 
field model~\cite{He2006_EPJA28-265}. 
Later the SS in anti-particle spectrum was studied with 
the relativistic Hartree-Fock model and the contribution from the Fock
term was discussed~\cite{Liang2010_EPJA44-119}.
It has been pointed out in Ref.~\cite{Zhou2003_PRL91-262501} that one open problem 
related to the study of the SS in anti-nucleon spectra is the polarization effect
caused by the annihilation of an anti-nucleon in a normal nucleus.
Detailed calculations of the antibaryon ($\bar p$, $\bar\Lambda$, etc.) annihilation 
rates in the nuclear environment showed that the in-medium annihilation rates are 
strongly suppressed by a significant reduction of the reaction $Q$ values, 
leading to relatively long-lived antibaryon-nucleus systems~\cite{Mishustin2005_PRC71-035201}.
Recently the SS in the anti-$\Lambda$ spectrum of hypernuclei was studied 
quantitatively~\cite{Song2009_CPL26-122102,*Song2011_CPL28-092101,*Song2010_ChinPhysC34-1425};
this kind of study would be of great interests for possible experimental tests.

The SS and PSS have been investigated extensively within the relativistic framework.
The readers are referred to Ref.~\cite{Ginocchio2005_PR414-165} 
for a review and to Ref.~\cite{Liang2013_PRC87-014334} for an overview of recent progresses.
Next we briefly mention several aspects of these progresses.
The node structure of radial Dirac wavefunctions of pseudospin 
doublets was studied in Ref.~\cite{Leviatan2001_PLB518-214} which was
helpful particularly for the understanding of 
the special status of nodeless intruder states in nuclei.
Although there are some doubts about 
the connection between the PSS and conditions
$\Sigma(r)=0$ or $d\Sigma(r)/dr=0$~\cite{Marcos2007_EPJA34-429,
*Marcos2008_EPJA37-251,*Marcos2008_JPCS128-012034,*Desplanques2010_EPJA43-369},
following the idea that under these conditions
the PSS is conserved exactly,
a lot of discussions have been made about the PSS and/or SS
in single (anti-)particle spectra obtained by exactly or approximately
solving the Dirac Hamiltonian with various potentials~\cite{Jia2007_EPJA34-41,
Zhang2008_PRA78-040101R,*Zhang2009_PRA80-054102,*Zhang2009_JMP50-032301,
Zhang2009_PS80-065018,*Zhang2009_CEJP7-768,*Zhang2009_IJTP48-2625,*Zhang2011_JMP52-053518, 
Setare2009_APPB40-2809,*Setare2010_APPB41-2459,*Setare2010_MPLA25-549,
Wei2010_PLB686-288, *Wei2010_EPJA43-185, *Chen2011_ChinPhysB20-062101,
Aydogdu2010_EPJA43-73,*Aydogdu2011_PLB703-379,*Aydogdu2013_ChinPhysB22-010302,
Akcay2009_PLA373-616,*Akcay2009_IJMPC20-930, 
Hamzavi2010_PLA374-4303,*Ikhdair2012_FBS53-473,*Hamzavi2012_IJMPE21-1250097,
*Hamzavi2013_AdvHEP2013-196986, 
Zarrinkamar2011_IJMPA26-1011, *Hassanabadi2012_ChinPhysB21-120302, *Maghsoodi2012_PS85-055007,
Candemir2012_IJMPE21-1250060, Wang2013_JPG40-045105,
Falaye2013_ChinPhysB22-060305}.
One of interesting topics is the tensor effect on the PSS or SS
which have been investigated in some of the above mentioned work and 
some others, e.g., in Refs.~\cite{Lisboa2004_PRC69-024319,
*Alberto2005_PRC71-034313,*Castro2006_PRC73-054309,*Castro2012_PRC86-052201,
Long2007_PRC76-034314, Long2010_PRC81-031302R,
Typel2008_NPA806-156}.
Much efforts were also devoted to the study of the perturbative feature of
the breaking of the PSS~\cite{Lisboa2010_PRC81-064324,*Castro2012_PRA86-032122,
Liang2011_PRC83-041301R,*Li2011_ChinPhysC35-825, Liang2013_PRC87-014334}.
The concept of supersymmetry (SUSY) and the similarity renormalization group method
were both used in the study of the PSS and/or SS by
several groups~\cite{Leviatan2009_PRL103-042502, Guo2012_PRC85-021302R, *Li2013_PRC87-044311,
Liang2013_PRC87-014334}.
The relevance of the PSS in the structure of halo 
nuclei~\cite{Long2010_PRC81-031302R} and superheavy nuclei~\cite{Jolos2007_PAN70-812, 
Li2013_arXiv1303.2765} was also found.
Quite recently, the node structure of radial Dirac wavefunctions in central 
confining potentials was studied and the authors have shown in a general way that 
it is possible to have positive energy bound solutions for these potentials
under the condition of exact PSS~\cite{Alberto2013_PRC87-031301R}.
Note that there have been some investigations of PSS connected 
with some specific forms of confining 
potentials~\cite{Chen2003_CPL20-358,*Chen2003_HEPNP27-324}.
Finally we mention that there have been some discussions on the physics 
behind $\Sigma(r) \approx 0$ or $d\Sigma(r)/dr \approx 0$~\cite{Furnstahl2000_NPA673-298, 
Ke2010_IJMPE25-1123} and more investigations are expected.  

In recent years, there has been an increasing interest in the exploration 
of continuum and resonant states especially in the study of exotic nuclei with 
unusual $N/Z$ ratios.
In these nuclei, the neutron (or proton) Fermi surface is close to the particle continuum, 
thus the contribution of the continuum is 
important~\cite{Dobaczewski1984_NPA422-103,*Dobaczewski1996_PRC53-2809, 
Meng1996_PRL77-3963,*Meng1998_PRL80-460,*Meng1998_NPA635-3,
*Meng2002_PRC65-041302R,*Meng2006_PPNP57-470,*Meng2011_SciChinaPMA54S1-119,
Poschl1997_PRL79-3841,*Vretenar2005_PR409-101,
Zhou2010_PRC82-011301R,*Li2012_PRC85-024312,*Li2012_CPL29-042101,*Li2012_AIPCP1491-208,
Zhang2011_PRC83-054301,*Zhang2012_PRC86-054318,
Pei2008_PRC78-064306,*Pei2011_PRC84-024311,*Pei2013_PRC87-051302R,
He2011_SciChinaPMA54S1-32,
Lin2011_SciChinaPMA54S1-73,
Lu2011_SciChinaPMA54S1-136}. 
Many methods or models developed for the study of resonances~\cite{Kukulin1989}
have been adopted to locate the position 
and to calculate the width of a nuclear resonant state,
e.g., the analytical continuation in coupling constant (ACCC) 
method~\cite{Yang2001_CPL18-196, Zhang2004_PRC70-034308,
*Zhang2007_EPJA32-43,*Zhang2009_IJMPE18-1761,*Zhang2012_PRC86-032802,*Zhang2012_EPJA48-40},
the real stabilization method (RSM)~\cite{Zhang2008_PRC77-014312,
*Zhang2007_APS56-3839,*Zhou2009_JPB42-245001, Mei2009_ChinPC33S1-101,
Zhang2009_ChinPC33-187,*Zhang2010_MPLA25-727},
the complex scaling method (CSM)~\cite{Guo2010_PRC82-034318,*Guo2010_IJMPE19-1357,
*Guo2010_CPC181-550,*Liu2012_PRC86-054312},
the coupled-channel method~\cite{Hagino2004_NPA735-55, Li2010_PRC81-034311,*Li2010_SCG53-773},
and some others~\cite{Fedorov2009_FBS45-191, Fernandez2012_AppMathComp218-5961}.
Each of these methods has advantages and disadvantages.
For example, the RSM is very powerful for narrow resonances and
the CSM for broad ones.

The study of symmetries in resonant states is certainly interesting; 
one of the topics is the PSS in the continuum.
We note that the PSS and/or SS in nucleon-nucleus and nucleon-nucleon scatterings have
been investigated~\cite{Ginocchio1999_PRL82-4599,*Ginocchio2002_PRC65-054002, 
Leeb2000_PRC62-024602,*Leeb2004_PRC69-054608}.
Meanwhile, there were also some numerical investigations of the PSS in single particle 
resonances~\cite{Guo2005_PRC72-054319,*Guo2006_PRC74-024320,*Liu2013_PRA87-052122, 
Zhang2006_HEPNP30S2-97,*Zhang2007_CPL24-1199} and
the SS in single particle resonant states~\cite{Xu2012_IJMPEE21-1250096}.
Recently, we gave a rigorous justification of the PSS in single particle 
resonant states~\cite{Lu2012_PRL109-072501}.
We have shown that the PSS in single particle resonant states in nuclei
is exactly conserved under the same condition for the PSS in bound states.
i.e., $\Sigma(r)=0$ or $d\Sigma(r)/dr=0$~\cite{Lu2012_PRL109-072501, Lu2013_AIPCP1533-63}. 
As we noted in Ref.~\cite{Lu2012_PRL109-072501}, 
it is straightforward to extend the study of the PSS 
in resonant states in the Fermi sea to that in the negative energy states 
in the Dirac sea or SS in anti-particle continuum spectra.
In the present work we will focus on several open problems related to the exact 
conservation and breaking mechanism of the PSS in single particle resonances.
To this end spherical square well potentials are employed
in which the PSS breaking part can be separated from other parts in the Jost function.
By examining zeros of Jost functions corresponding 
to small components of radial Dirac wave functions, 
we examine general properties of PSS splittings of the energies and widths.

In Sec.~\ref{sec:formalism}, the justification of the PSS in single
particle resonant states will be given briefly and the emphasis will be put on the
mechanism of the exact conservation and breaking of pseudospin symmetry in single 
particle resonant states in square well potentials. 
Details on the study of the PSS in single particle resonances 
will be presented in Sec.~\ref{sec:results}.
In Sec.~\ref{sec:summary}, we will summarize this work and mention some
perspectives.

\section{\label{sec:formalism}
Pseudospin symmetry in single 
particle resonant states in square well potentials}

A rigorous justification of the PSS in single particle resonant states 
has been given in Ref.~\cite{Lu2012_PRL109-072501}.
For completeness, we will briefly mention it before we discuss 
the PSS in square well potentials. 

In relativistic mean field models, the covariant functional can be 
one of the following four forms: the meson exchange or point-coupling nucleon 
interactions combined with the non-linear or density-dependent 
couplings~\cite{Serot1986_ANP16-1, Reinhard1989_RPP52-439, 
Ring1996_PPNP37-193, *Ring2001_PPNP46-165,
Vretenar2005_PR409-101, Meng2006_PPNP57-470, Paar2007_RPP70-691, Meng2013_FPC8-55}.
The starting point of the covariant density functional with the non-linear
point-couplings is the following Lagrangian:
\begin{equation}
 \mathcal{L} = \bar{\psi}(i\gamma_{\mu}\partial^{\mu}-M)\psi 
              -\mathcal{L}_{{\rm lin}}
              -\mathcal{L}_{{\rm nl}}
              -\mathcal{L}_{{\rm der}}
              -\mathcal{L}_{{\rm Cou}},
\end{equation}
where
\begin{eqnarray}
 \mathcal{L}_{{\rm lin}} & = & \frac{1}{2} \alpha_{S} \rho_{S}^{2}
                              +\frac{1}{2} \alpha_{V} (j_{V})_\mu j^\mu_{V}
 \nonumber \\
 &  & \mbox{}
                              +\frac{1}{2} \alpha_{TS} (\vec{\rho}_{TS})^{2}
                              +\frac{1}{2} \alpha_{TV} (\vec{j}_{TV})_\mu \vec{j}^\mu_{TV},
 \nonumber \\
 \mathcal{L}_{{\rm nl}}  & = & \frac{1}{3} \beta_{S} \rho_{S}^{3}
                              +\frac{1}{4} \gamma_{S}\rho_{S}^{4} 
                              +\frac{1}{4} \gamma_{V}[(j_{V})_\mu j^\mu_{V}]^{2} ,
 \nonumber \\
 \mathcal{L}_{{\rm der}} & = & \frac{1}{2} \delta_{S} \partial_{\nu}\rho_{S}     
                                                      \partial^{\nu}\rho_{S}     
                              +\frac{1}{2} \delta_{V} \partial_{\nu}(j_{V})_{\mu} 
                                                      \partial^{\nu} j_{V}^{\mu} 
 \nonumber \\
 &  & \mbox{}                 
                              +\frac{1}{2} \delta_{TS} \partial_{\nu}\vec{\rho}_{TS} 
                                                       \partial^{\nu}\vec{\rho}_{TS} 
                              +\frac{1}{2} \delta_{TV} \partial_{\nu} (\vec{j}_{TV})_\mu 
                                                       \partial^{\nu}  \vec{j}^\mu_{TV} ,
 \nonumber \\
 \mathcal{L}_{{\rm Cou}} & = & \frac{1}{4} F^{\mu\nu} F_{\mu\nu}
                             +e\frac{1-\tau_{3}}{2} A_{\mu} j_{V}^{\mu} ,
\end{eqnarray}
are the linear coupling, non-linear coupling, derivative coupling, 
and the Coulomb part, respectively.
$M$ is the nucleon mass, $\alpha_{S}$, $\alpha_{V}$, $\alpha_{TS}$,
$\alpha_{TV}$, $\beta_{S}$, $\gamma_{S}$, $\gamma_{V}$, $\delta_{S}$, 
$\delta_{V}$, $\delta_{TS}$, and $\delta_{TV}$ are coupling constants 
for different channels and $e$ is the electric charge.
$\rho_{S}$, $\vec{\rho}_{TS}$, $j_{V}$, and $\vec{j}_{TV}$ are the iso-scalar density, 
iso-vector density, iso-scalar current, and iso-vector current, respectively. 
The various densities and currents are defined as:
\begin{eqnarray}
      \rho  _{S} = \bar{\psi}\psi , & \qquad & 
 \vec{\rho}_{TS} = \bar{\psi}\vec{\tau}\psi ,
 \label{eq:densities}
 \\
      j _{V}^{\mu} = \bar{\psi} \gamma^{\mu} \psi , & \qquad & 
 \vec{j}_{TV}^{\mu} = \bar{\psi} \vec{\tau} \gamma^{\mu} \psi.
 \label{eq:currents}
\end{eqnarray}

The equation of motion for nucleons, the Dirac equation, is derived from
the Lagrangian density as,
\begin{equation}
 \left[ \bm{\alpha}\cdot\bm{p} + \beta(M+S(\bm{r}))+V(\bm{r}) \right]\psi(\bm{r})
 = \epsilon \psi(\bm{r}),
~\label{Eq:Diraceq}
\end{equation}
where $\bm{\alpha}$ and $\beta$ are Dirac matrices and $M$
is the nucleon mass. 
In Eq.~(\ref{Eq:Diraceq}), the scalar potential $S(\bm{r})$ and the vector potential 
$V(\bm{r})$ are determined by densities and currents defined in Eqs.~(\ref{eq:densities})
and (\ref{eq:currents}). 
It turns out that both potentials are very deep, but they have opposite signs:
The scalar potential $S(\bm{r})$ is attractive and the vector potential $V(\bm{r})$
is repulsive. This results in an approximate PSS in nuclear single particle 
spectra~\cite{Ginocchio1997_PRL78-436}
and an even better conserved SS in anti-nucleon spectra~\cite{Zhou2003_PRL91-262501}. 

For a spherical nucleus, the Dirac spinor reads,
\begin{equation}
\psi(\bm{r})=\frac{1}{r}\left(\begin{array}{c}
iF_{n\kappa}(r)Y_{jm}^{l}(\theta,\phi)\\
-G_{\tilde{n}\kappa}(r)Y_{jm}^{\tilde{l}}(\theta,\phi)
\end{array}\right),
\end{equation}
where $Y_{jm}^{l}(\theta,\phi)$ is the spin spherical harmonic,
$F_{n\kappa}(r)/r$ and $G_{\tilde{n}\kappa}(r)/r$ are the radial wave functions for
the upper and lower components with $n$ and $\tilde{n}$ numbers of radial nodes. 
The total angular momentum $j$, the orbit angular momentum $l$, and the pseudo orbital
angular momentum $\tilde{l}$ are determined by $\kappa$ through
\begin{eqnarray}
                        j & = & |\kappa|-\frac{1}{2},  \nonumber \\
                   l(l+1) & = & \kappa(\kappa+1),   \quad            l\geq0, \nonumber\\
   \tilde{l}(\tilde{l}+1) & = & \kappa(\kappa-1),   \quad    \tilde{l}\geq0.
~\label{Eq:angmom}
\end{eqnarray}
The radial Dirac equation reads
\begin{equation}
\left(\begin{array}{cc}
 M + \Sigma(r)                 & -\frac{d}{dr}+\frac{\kappa}{r} \\
 \frac{d}{dr}+\frac{\kappa}{r} & -M+\Delta(r)
\end{array}\right)
\left(\begin{array}{c}
 F(r) \\
 G(r)
\end{array}\right)
= \epsilon
\left(\begin{array}{c}
 F(r) \\
 G(r)
\end{array}\right),
\label{Eq:Diracradial}
\end{equation}
where $\Sigma(r)=V(r)+S(r)$, $\Delta(r)=V(r)-S(r)$, and $\epsilon$ is the eigenenergy. 
For brevity we omit the subscripts from $F(r)$ and $G(r)$. 
This first order coupled equation can be rewritten as two decoupled
second order differential ones by eliminating either the large or
the small component~\cite{Meng1998_PRC58-R628, Meng1999_PRC59-154, Zhou2003_PRL91-262501},
\begin{eqnarray}
 \left[
   \frac{d^{2}}{dr^{2}}
  +\frac{1}{M_{+}(r)}\dfrac{d\Delta(r)}{dr}\frac{d}{dr}
  -\dfrac{{l}({l}+1)}{r^{2}}
  \hspace*{1cm}
 \right.
 &   &
 \nonumber \\
 \left. \mbox{}
  + \dfrac{1}{M_{+}(r)}
    \frac{\kappa}{r}\dfrac{d\Delta(r)}{dr}
  -  M_+(r) M_-(r)
 \right] F(r)
 & = & 
 0 , 
 \label{eq:F}
 \\
 \left[
   \frac{d^{2}}{dr^{2}}
  -\frac{1}{M_{-}(r)}\dfrac{d\Sigma(r)}{dr}\frac{d}{dr}
  -\dfrac{\tilde{l}(\tilde{l}+1)}{r^{2}}
  \hspace*{1cm}
 \right.
 &   &
 \nonumber \\
 \left. \mbox{}
  + \dfrac{1}{M_{-}(r)}
    \frac{\kappa}{r}\dfrac{d\Sigma(r)}{dr}
  -  M_+(r) M_-(r)
 \right] G(r)
 & = & 
 0 , 
 \label{eq:G}
\end{eqnarray}
where $M_+(r)=M+\epsilon-\Delta(r)$ and $M_-(r)=M-\epsilon+\Sigma(r)$.
In Ref.~\cite{Zhang2010_IJMPE19-55,*Zhang2009_ChinPC33S1-113,
*Zhang2009_CPL26-092401,*Li2011_SciChinaPMA54-231},
it has been shown that each of these two Schr\"odinger-like equations, 
together with its charge conjugated one, are fully equivalent to Eq.~(\ref{Eq:Diraceq}).
Note that for bound states, there is always a singularity in $1/M_{-}(r)$ in Eq.~(\ref{eq:G}). 
For resonant states we discuss here, such a singularity does not exist.

For the continuum in the Fermi sea, i.e., $\epsilon\geq M,$ there
exist two independent solutions for Eqs.~(\ref{eq:F}) or (\ref{eq:G}). 
The physically acceptable one is the solution that vanishes at the origin.
For example, the regular solution for the small component $G(r)$
behaves like $j_{\tilde{l}} (pr)$ as $r\rightarrow0$,
\begin{equation}
 \lim_{r\rightarrow0}G(r)/j_{\tilde{l}} (pr)=1,\qquad p=\sqrt{\epsilon^{2}-M^{2}}.
~\label{eq:Gatorigin}
\end{equation}

Since the PSS is directly connected with the small component,
we will mainly focus on Eq.~(\ref{eq:G}) in the following discussions.
At large $r$ potentials for neutrons vanish,
the regular solution is written as a combination of Riccati-Hankel functions,
\begin{equation}
 G(r) = \frac{i}{2}\left[\mathcal{J}^G_{\kappa}(p)     h_{\tilde{l}}^{-}(pr) - 
                         \mathcal{J}^G_{\kappa}(p)^{*} h_{\tilde{l}}^{+}(pr) \right],
 \ \ r\rightarrow\infty,
\label{Eq:Jost_Gdef}
\end{equation}
where $h_{\tilde{l}}^{\pm}(pr)$ the Ricatti-Hankel functions.
$\mathcal{J}_{\kappa}(p)$ is the Jost function 
which is an analytic function of $p$ and can be analytically continued
to a large region in the complex $p$ plane. 
The zeros of $\mathcal{J}_{\kappa}^{G}(p)$ on the positive imaginary axis of the $p$ plane
correspond to bound states and those on the lower $p$ plane and 
near the real axis correspond to resonant states. 
The resonance energy $E$ and width $\Gamma$ are determined by the relation 
$E-i{\Gamma}/{2}=\epsilon - M$. 

In the PSS limit, Eq.~(\ref{eq:G}) is reduced as
\begin{equation}
 \left[
          \frac{d^{2}}{dr^{2}}
        - \frac{\tilde{l}(\tilde{l}+1)}{r^{2}}
        + \left(
                \epsilon-M
          \right)
          M_+(r)
 \right]  G(r)
 =  0
 .
~\label{eq:PSLeqG}
\end{equation}
For pseudospin doublets with $\kappa$ and 
$\kappa^{\prime}=-\kappa+1$, the small components satisfy the same equation because 
they have the same pseudo orbital angular momentum $\tilde{l}$.
For continuum states, $G_{\kappa}(\epsilon,r)=
G_{\kappa^{\prime}}(\epsilon,r)$ is true for any energy $\epsilon$ thus 
we have $\mathcal{J}_{\kappa^{\prime}}^{G}(p)=\mathcal{J}_{\kappa}^{G}(p)$. 
This equivalence can be generalized into the complex $p$ plane 
due to the uniqueness of the analytic continuation. 
Thus the zeros are the same for $\mathcal{J}_{\kappa^{\prime}}^{G}(p)$ and 
$\mathcal{J}_{\kappa}^{G}(p)$ and the PSS in single particle resonant states 
in nuclei is exactly conserved.
Note that when we focus on the zeros of Jost functions of pseudospin doublets
on the positive imaginary axis of the $p$ plane, we come to the well-known PSS for bound states.
For bound states this is an alternative way to justify the PSS or SS
besides those ways in which the Dirac equation and Dirac wave functions 
are examined in, e.g., Refs.~\cite{Ginocchio1997_PRL78-436,
Meng1998_PRC58-R628, Lalazissis1998_PRC58-R45, Zhou2003_PRL91-262501} 
or introducing the concept of SUSY and the similarity renormalization group 
method~\cite{Leviatan2009_PRL103-042502, Guo2012_PRC85-021302R, *Li2013_PRC87-044311,
Liang2013_PRC87-014334}.
For single particle resonant states, up to now this is the only way to do so.

\begin{figure}
\begin{centering}
\includegraphics[width=0.8\columnwidth]{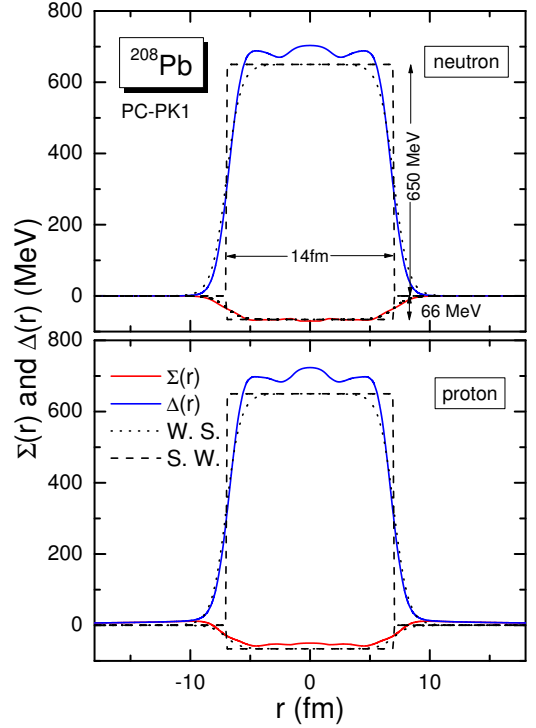}
\par\end{centering}
\caption{\label{potential}(Color online).
Potentials $\Sigma(r)$ and $\Delta(r)$ of $^{208}$Pb calculated 
using the relativistic mean field model with the parameter set 
PC-PK1~\cite{Zhao2010_PRC82-054319,*Zhao2011_PRL107-122501}.
These potentials can be approximated by Woods-Saxon (W.S.)
potentials (dotted curves) or spherical square well (S.W.) potentials (dashed lines)
with a radius around 7 fm and depths around 650 MeV and 66 MeV. 
The long tails in the proton case are due to the Coulomb interaction.
}
\end{figure}

In Fig.~\ref{potential} we show spherical potentials $\Sigma(r)$
and $\Delta(r)$ of $^{208}$Pb calculated using the relativistic mean
field model with the parameter set 
PC-PK1~\cite{Zhao2010_PRC82-054319,*Zhao2011_PRL107-122501}.
However, to extract the energy and width of resonant states in such potentials 
is relatively complex.
In particular, when we want to study the PSS and examine the origin and 
the splitting mechanism, it is better to start from solvable models. 
In Fig.~\ref{potential} we show two types of potentials which
can be used to approximate the realistic one, the Woods-Saxon and
square well potentials. 
For $^{208}$Pb the radius is around 7 fm and the depths of potentials 
are 650 MeV and 66 MeV, respectively; 
these parameters of Woods-Saxon potentials 
have been proposed in Ref.~\cite{Guo2005_PRC72-054319}.
Although the diffuseness of realistic potentials can not be included,
it is still a good starting point to study general properties of the PSS 
for the resonant as well as bound states by using square well potentials 
because the PSS-breaking term in the Jost function is separated from the PSS-conserving term. 

Spherical square well potentials for $\Sigma(r)$ and $\Delta(r)$ read,
\begin{eqnarray}
 \Sigma(r) & = & \left\{ \begin{array}{c}
                         C,\qquad r<R,\\
                         0,\qquad r\geq R,
                         \end{array}\right.\\
 \Delta(r) & = & \left\{ \begin{array}{c}
                         D,\qquad r<R,\\
                         0,\qquad r\geq R,
                         \end{array}\right.
 \label{eq:S.W.}
\end{eqnarray}
where $C$ and $D$ are depths and $R$ is the width. 
The Jost function $\mathcal{J}_{\kappa}^{G}(p)$ is derived as~\cite{Lu2012_PRL109-072501},
\begin{eqnarray}
 \mathcal{J}_{\kappa}^{G}(p) & = & 
 - \frac{p^{\tilde{l}}}{2ik^{\tilde{l}+1}}
   \left\{ j_{\tilde{l}}(kR) p h_{\tilde{l}}^{+\prime}(pR)
        -kj_{\tilde{l}}^{\prime}(kR) h_{\tilde{l}}^{+}(pR)
    \phantom{\frac{C}{C}}
   \right.
   \nonumber \\
 &  & 
   \left. \mbox{}
   - \frac{C}{\epsilon-M-C}
     \left[ kj_{\tilde{l}}^{\prime}(kR) - \frac{\kappa}{R} j_{\tilde{l}}(kR) \right] 
     h_{\tilde{l}}^{+}(pR)
   \right\}, 
   \nonumber \\ 
 \label{eq:Jost_function}
\end{eqnarray}
with $k = \sqrt{\left( \epsilon - C - M \right) \left( \epsilon - D + M \right)}$.
The PSS in both bound states and resonant states can be explained explicitly. 
If $C = 0$, 
the second term in $\mathcal{J}_{\kappa}^{G}(p)$ vanishes and 
the first term only depends on the pseudo orbital angular momentum $\tilde{l}$. 
Then Jost functions with different $\kappa$ but the same $\tilde{l}$ 
are identical, energies 
and widths of resonant pseudospin partners are exactly the same.

For the large component $F(r)$ we can write down a similar expression for the
asymptotic behavior,
\begin{equation}
 F(r) = \frac{i}{2} 
        \left[ \mathcal{J}_{\kappa}^{F}(p)     h_{{l}}^{-}(pr) - 
               \mathcal{J}_{\kappa}^{F}(p)^{*} h_{{l}}^{+}(pr)
        \right],
 \ \ r\rightarrow\infty.
\end{equation}
At the origin,
\begin{equation}
 \lim_{r\rightarrow0} F(r) / j_{l}(pr) = 1,
 \ \ p=\sqrt{\epsilon^{2}-M^{2}}.
\end{equation}
In square well potentials, $\mathcal{J}_{\kappa}^{F}(p)$ is derived as,
\begin{eqnarray}
 \mathcal{J}_{\kappa}^{F}(p) 
 & = & 
 -\frac{p^{l}}{2ik^{l+1}}
  \left\{ j_{l}(kR) p h_{l}^{+\prime}(pR) - k j_{l}^{\prime}(kR) h_{l}^{+}(pR)
   \phantom{\frac{D}{D}}
  \right. \nonumber\\
 &   & 
  \left. \mbox{}
  -\frac{D}{\epsilon+M-D} 
   \left[ k j_{l}^{\prime}(kR) + \frac{\kappa}{R} j_{l}(kR) \right] h_{l}^{+}(pR)
  \right\}. \nonumber\\
\end{eqnarray}
It looks similar to that of the small component $\mathcal{J}_{\kappa}^{G}(p)$,
with the exception that the potential parameter $C$ is substituted by 
$D$ and the pseudo orbital angular momentum $\tilde{l}$
is substituted by $l$. 
In the case of $D\rightarrow0$, this form of Jost function can be used to 
investigate the spin symmetry of single particle levels.

By examining the zeros of the Jost function, we will study properties
of single particle bound states and resonant states and the PSS.

\section{\label{sec:results}Results and discussions}

\subsection{The Jost function and its zeros and the occurrence of intruder states}

\begin{figure}
\begin{centering}
\includegraphics[width=0.98\columnwidth]{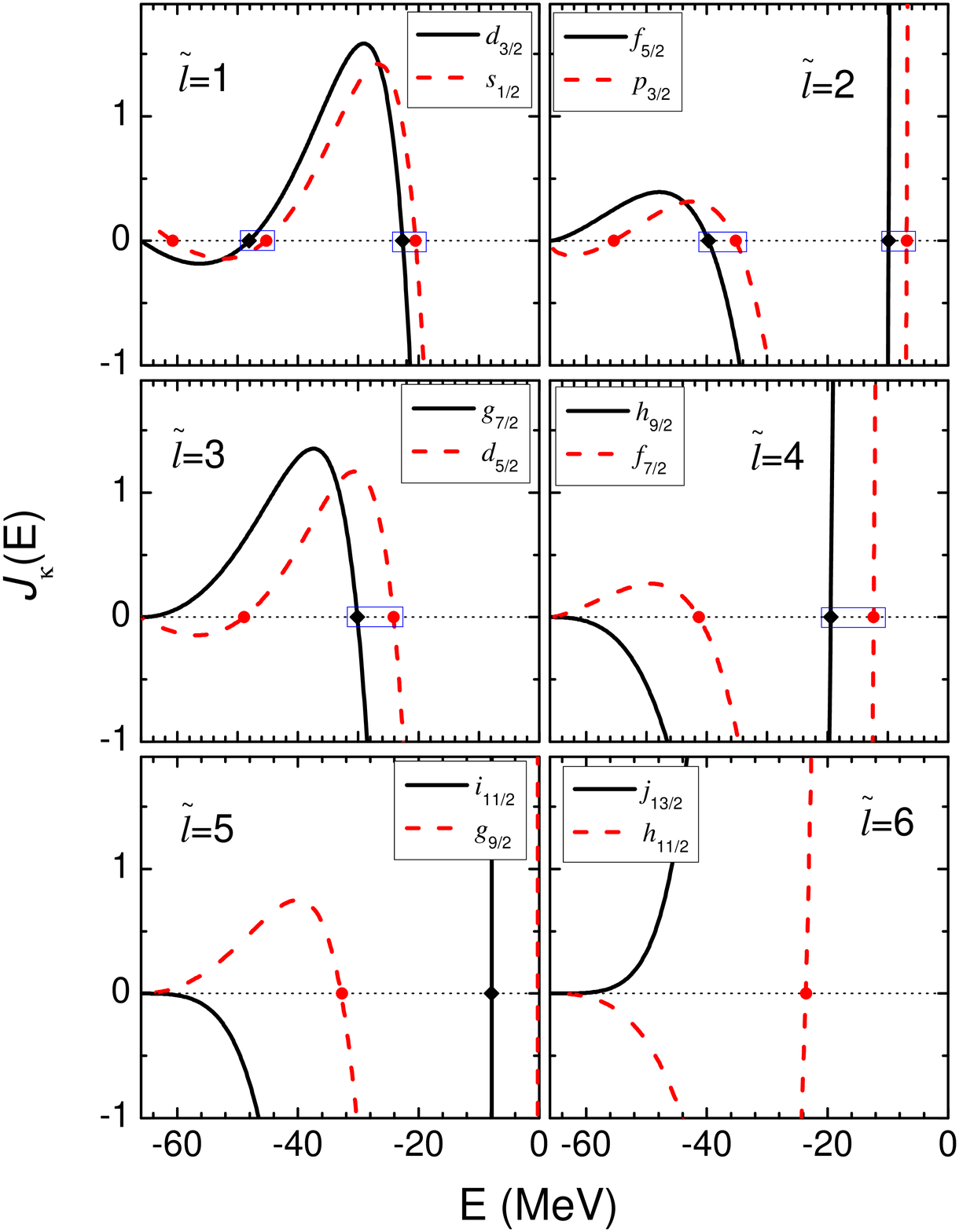}
\par\end{centering}
\caption{\label{fig:comparison_of_Jost_functions}(Color online).
The Jost function $\mathcal{J}_\kappa(E)$ (in arbitrary unit) 
on the real $E$ axis for several pairs of pseudospin partners. 
The results for pseudospin $\tilde{s}=\pm 1/2$ 
are denoted as solid and dashed curves, respectively. 
The zero points representing bound states with $\tilde{s}=\pm{1}/{2}$ 
are denoted as black and red dots, respectively. 
}
\end{figure}

Now we examine the Jost function (\ref{eq:Jost_function}) corresponding to
the small component of the radial Dirac wavefunction in square well potentials
with $C=-66$~MeV and $D=650$~MeV.
For bound states, one can draw the Jost function as a function of
either the imaginary part of $p$ or the binding energy $E\equiv\epsilon-M$. 
Here we use the latter in order to have a more intuitive picture for the energy splitting.
Such Jost functions $\mathcal{J}_{\kappa}(E)$ for several pairs of pseudospin partners
are shown in Fig.~\ref{fig:comparison_of_Jost_functions}.
In the following discussions, we omit the superscript ``G'' from $\mathcal{J}_{\kappa}(E)$.
The results for pseudospin $\tilde{s}=\pm 1/2$ 
are denoted as solid and dashed curves, respectively. 
The zero points representing bound states with $\tilde{s}=\pm 1/2$ 
are denoted as black and red dots, respectively. 
For each pseudo orbital angular momentum $\tilde{l}$ there exist two
$\kappa$'s, one with a positive value $\kappa = \tilde{l}+1$ and 
the other with a negative value $\kappa = -\tilde{l}$, respectively. 
For example, for $\tilde{l}=1$ ($\tilde{p}$) we have $\kappa=2$
(pseudo spin aligned, $d_{3/2}$) or $\kappa=-1$ (pseudo spin anti-aligned, $s_{1/2}$). 
There are some common features in these Jost functions which will
be detailed in the following.

First, the number of zeros for Jost functions with negative $\kappa$ is
always one more than that with positive $\kappa$ if the non-physical node
at the bottom of the potential is excluded.
For example, for $\tilde{l}=1$ there are two zeros for $d_{3/2}$ ($\kappa=2$) 
but three for $s_{1/2}$ ($\kappa=-1$); 
this means that there are two bound states for $d_{3/2}$ and three for $s_{1/2}$. 
Therefore, there is always one bound state with $\kappa<0$ which does not
have a partner; this state is simply an intruder state.
The study of the node structure of radial Dirac wavefunctions of pseudospin 
doublets has been made in which the occurrence of nodeless intruder states
has been explained~\cite{Leviatan2001_PLB518-214}.
This kind of study was also extended to the Dirac sea~\cite{Zhou2003_PRL91-262501}
and to the case with confining potentials~\cite{Chen2003_CPL20-358,*Chen2003_HEPNP27-324,
Alberto2013_PRC87-031301R}.
Note that for harmonic oscillator (HO) potentials or some combination of
HO and Woods-Saxon potentials, bound states exists 
even under the condition of the exact PSS and 
intruder states do have pseudospin partners~\cite{Chen2003_CPL20-358,*Chen2003_HEPNP27-324}.
The reason why there are intruder states was also naturally explained 
by employing both exact and broken SUSY within a unified scheme~\cite{Leviatan2004_PRL92-202501,
Leviatan2009_PRL103-042502, Liang2013_PRC87-014334}.
Here we present a novel way to show the origin of the appearance of intruder states:
The lowest zero of the Jost function (\ref{eq:Jost_function}) with negative $\kappa$ 
is always isolated while the others are paired off with those of the Jost function
with positive $\kappa$. 

Second, the similarity of Jost functions can also be used as a test
of the PSS or SS symmetries 
besides the examination of radial wave functions of PSS or SS 
doublets~\cite{Ginocchio1998_PRC57-1167,*Ginocchio2002_PRC66-064312,
Lalazissis1998_PRC58-R45, Sugawara-Tanabe2002_PRC65-054313, Ginocchio2004_PRC69-034303,
He2006_EPJA28-265, Zhang2006_HEPNP30S2-97,*Zhang2007_CPL24-1199, Typel2008_NPA806-156}.
From Fig.~\ref{fig:comparison_of_Jost_functions}, one finds that 
except in the energy range around and below the first zero
of the Jost function with negative $\kappa$,
Jost functions corresponding to PSS doublets, i.e., 
with the same $\tilde{l}$, are similar with each other. 
This is particularly clear for the cases with small $\tilde{l}$, e.g., $\tilde{l}=1$.
Furthermore, in the low energy range above the first zero of 
the Jost function with negative $\kappa$, these Jost functions are rather smooth.
When approaching the threshold, the Jost function becomes very steep against the energy. 
Even the absolute differences between $\mathcal{J}_{\kappa}$'s
are large, the zeros corresponding to PS doublets are still not far from each other,
which means a good PSS. 

Third, the difference between Jost functions with the same $\tilde{l}$ 
increases when $\tilde{l}$ becomes larger.
So is the difference of the positions of their paired zeros.
That is, the splitting between the energy (in short, the energy splitting) of pseudospin partners
increases with $\tilde{l}$ which is consistent with earlier studies, e.g,
in Ref.~\cite{Meng1999_PRC59-154, Lalazissis1998_PRC58-R45}.
This can be explained by the factor $\kappa$ of the pseudospin splitting term 
${\kappa} j_{\tilde{l}}(kR)/R$ in Eq.~(\ref{eq:Jost_function})
which breaks the PSS and
will be discussed in the next subsection in details.

\subsection{Single particle spectra and pseudospin splittings in the energy and the width}

\begin{figure}[t]
\begin{centering}
\includegraphics[width=0.9\columnwidth]{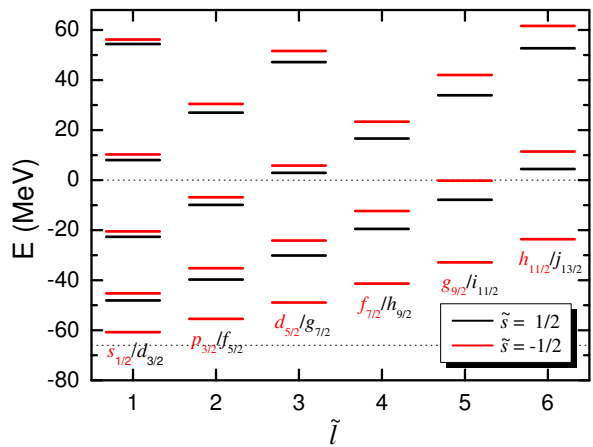}
\par\end{centering}
\caption{\label{fig:energies}(Color online).
Energies of bound states 
as well as resonances with $\tilde{l}=1, 2, \cdots, 6$ 
in spherical square well potentials with $C=-66$~MeV and $D=650$~MeV. 
The results with pseudospin $\tilde{s}=\pm{1}/{2}$
are denoted as black and red lines, respectively. 
The bottom of the potential well $E=-66$ MeV and the threshold $E=0$ 
are shown as dotted lines. 
For each pseudo orbital angular momentum $\tilde{l}$ the lowest level 
is an intruder state and has no pseudospin partner.
}
\end{figure}

There is no analytical solution for $\mathcal{J}_{\kappa}(p)=0$. 
However, because the Jost function is analytic near its zeros, 
one can easily get the roots of $\mathcal{J}_{\kappa}(p)=0$ by using 
the secant method. 
Starting from an initial guess for a root, the iteration usually converges 
after a few steps~\cite{Lu2012_PRL109-072501}. 

In Fig.~\ref{fig:energies} we show the energies of the bound states
as well as the resonances with $\tilde{l}=1, 2, \cdots, 6$ 
in square well potentials with $C=-66$~MeV and $D=650$~MeV.
The results with pseudospin $\tilde{s}=\pm{1}/{2}$
are denoted as black and red lines, respectively. 
To see PSS splittings more clearly, the results are depicted with
respect to the pseudo orbital angular momentum $\tilde{l}$. 
For example, for $\tilde{l}=3$ we have put together 
levels with $g_{7/2}$ ($\tilde{s}={1}/{2}$ and $\kappa=4$) and 
those with $d_{5/2}$ ($\tilde{s}=-{1}/{2}$ and $\kappa=-3$).


As explained in the previous subsection, 
for each pseudo orbital angular momentum $\tilde{l}$, the lowest level
represents an intruder state which has no pseudospin partner. 
In Fig.~\ref{fig:energies}, one finds the normal energy splitting,
i.e., for a pair of pseudospin doublet states, 
the one with $\tilde{s}=-1/2$ is higher in energy than that with $\tilde{s}=1/2$, 
regardless these states are bound or in the continuum.
For bound states, this is well known, though in some realistic calculations, 
e.g., Ref.~\cite{Meng1999_PRC59-154, Lalazissis1998_PRC58-R45, Alberto2002_PRC65-034307}, 
it is shown that there are some exceptions.
For resonant states, from Refs.~\cite{Guo2005_PRC72-054319,*Guo2006_PRC74-024320, 
Zhang2006_HEPNP30S2-97,*Zhang2007_CPL24-1199}, one finds that in many cases,
for a pair of pseudospin doublet states, 
the one with $\tilde{s}=-1/2$ is lower in energy.
Systematic studies have been carried out to investigate the parameter (the depth,
the radius, and the diffuseness) dependence of the PSS in resonances 
in Woods-Saxon potentials and the isospin dependence of the PSS in resonances 
in RMF potentials~\cite{Guo2005_PRC72-054319,*Guo2006_PRC74-024320}.
It was found that in these more realistic potentials, the energy splitting
of pseudospin doublets in continuum has a complicated dependence on the parameters
of the potential and on the ratio of neutron and proton 
numbers~\cite{Guo2005_PRC72-054319,*Guo2006_PRC74-024320}.

For bound states, when approaching the threshold, 
the energy splitting becomes smaller for all $\tilde{l}$ as shown in
Fig.~\ref{fig:energies}.
This means that the PSS becomes more conserved for orbitals which
are closer to the continuum which has been found and 
explained~\cite{Ginocchio1997_PRL78-436, Meng1998_PRC58-R628, 
Meng1999_PRC59-154, Lalazissis1998_PRC58-R45}.
In the present work, we can explain this point by examining 
the Jost function (\ref{eq:Jost_function}).
The pseudospin splitting term in Eq.~(\ref{eq:Jost_function}) is proportional to 
$C/(\epsilon-M-C)$ and the denominator $\epsilon-M-C$ means 
the relative energy with respect to the bottom of the single particle potential.
Apparently, for less bound states, 
the factor $C/\epsilon-M-C$ is smaller and 
the pseudospin splitting term in the Jost function
are less important, consequently the PSS becomes better.

As we noticed earlier, the pseudospin splitting term in Eq.~(\ref{eq:Jost_function}) 
depends not only on the binding energy, but also on $\tilde{l}$ and $\kappa$.
Thus the energy splitting becomes larger as the pseudo orbital angular momentum 
$\tilde{l}$ increases.
Especially, if we focus on the levels with the same number of radial nodes, the energy splitting
is monotonically increasing with $\tilde{l}$; 
this behavior has been observed for bound states in many publications~\cite{Meng1998_PRC58-R628,
Meng1999_PRC59-154, Lalazissis1998_PRC58-R45, Zhou2003_PRL91-262501}.

\begin{figure}[t]
\begin{centering}
\includegraphics[width=0.9\columnwidth]{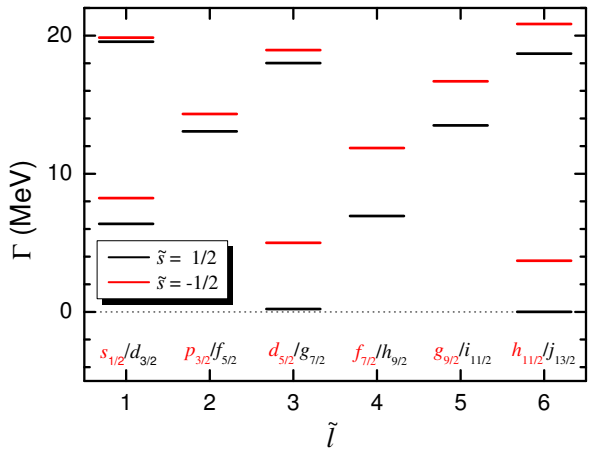}
\par\end{centering}
\caption{\label{fig:width}(Color online).
Widths of resonant states with $\tilde{l}=1, 2, \cdots, 6$ in spherical square well potentials
with $C=-66$~MeV and $D=650$~MeV. 
The results with pseudospin $\tilde{s}=\pm{1}/{2}$
are denoted as black and red lines, respectively. 
The bound state threshold $E=0$ is shown as dotted line.
}
\end{figure}

For resonant states, not only the energies but also the widths are of importance. 
In Fig.~\ref{fig:width} we show the calculated
widths of resonances in spherical square well potentials with $C=-66$ MeV and $D=650$ MeV. 
Comparing this figure with Fig.~\ref{fig:energies} one finds that 
the width splitting shares several similar features with the energy splitting.
First, the width of the resonant state with $\tilde{s}=-{1}/{2}$ is always
larger than that of its pseudospin partner with $\tilde{s}={1}/{2}$.
Note that the width splitting depends on the depth the potential $C$, 
as will be shown in Section~\ref{sec:dependence}.
Second, the width splitting decreases when the energy of resonant states
increases for the same $\tilde{l}$.
For the resonances with very high energies, the width splitting even becomes negligible.

From Fig.~\ref{fig:energies}, one finds that for resonant states, 
when the energy increases, the energy splitting becomes smaller for $\tilde{l} = 1$.
But it becomes a bit larger for $\tilde{l}=3$ and 5
which seems to contradict with the fact that the factor of the pseudospin splitting term 
in Eq.~(\ref{eq:Jost_function}), i.e., $C/(\epsilon-M-C)$ becomes smaller
in amplitude when the $\epsilon$ becomes larger. 
However, this is not really a contradiction because for a resonant state, one should
take into account both the energy and the width. 
For the closeness of pseudospin partners in the continuum, one should check
the distance between them in the complex energy or momentum plane.
As shown in Fig.~\ref{fig:dependency_on_potential_depth2}, for $\tilde{l}=2$ or 3, 
the distance between the $n$-th pair of pseudospin resonant doublets is
always smaller than that between the $(n-1)$-th pair when the potential depth is fixed.
Note that in Fig.~\ref{fig:dependency_on_potential_depth2} are presented 
bound states and resonant states in spherical square well potentials 
with $C=-70, -60, \cdots, 0$ MeV.

Different from the energy splitting which increases with $\tilde{l}$ increasing, 
from Fig.~\ref{fig:width} it is seen that when the radial quantum number is
fixed, the width splitting decreases with $\tilde{l}$
which seems inconsistent with the dependence of the pseudospin splitting term 
in Eq.~(\ref{eq:Jost_function}) on $\tilde{l}$ and $\kappa$.
To solve this puzzle, one again needs to examine the distance between PS
partners in the complex energy or momentum plane.
In Fig.~\ref{fig:dependency_on_potential_depth2}, one can find that,
at fixed $C$ and radial quantum number, e.g., comparing red curves in two subfigures,
the pair of $\tilde{l}=2$ pseudospin resonant partners is always closer than those of
$\tilde{l}=3$.

\subsection{\label{sec:dependence}
The dependence of the PSS in resonances on the depth of the potential:
An threshold effect in the energy splitting and an anomaly in the width splitting}

\begin{figure}
\begin{centering}
\includegraphics[width=0.95\columnwidth]{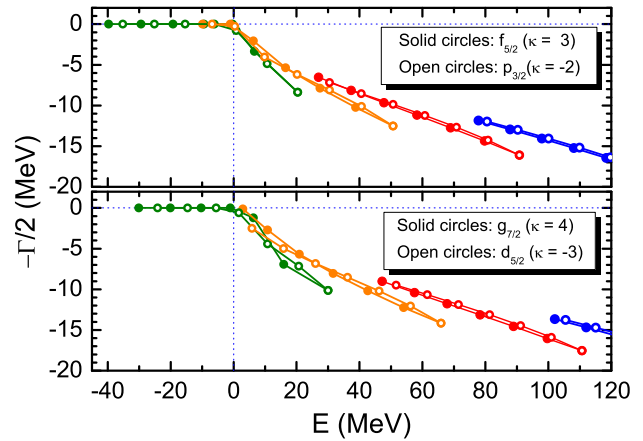}
\par\end{centering}
\caption{\label{fig:dependency_on_potential_depth2}(Color online).
Zeros of the Jost function $\mathcal{J}_{\kappa}(E)$ in spherical square well potentials 
with different potential depth $C=-70, -60, \cdots, 0$ MeV
for (a) $\tilde{l}=2$: $p_{3/2}$ ($\kappa=-2$) and $f_{5/2}$ ($\kappa=3$) 
and (b) $\tilde{l}=3$: $d_{5/2}$ ($\kappa=-3$) and $g_{7/2}$ ($\kappa=4$).
When $C=0$, pseudospin doublets are degenerate and corresponding symbols
overlap with each other.
The results with pseudospin $\tilde{s}=\pm{1}/{2}$ are denoted as solid and open
circles, respectively. 
When $C=-70$ MeV, energies of intruder states below $-40$ MeV and not shown.
}
\end{figure}

Since the potential depth is relevant mostly to the energy and width of a
resonant state, next we study how the PSS evolves with the variation of
the potential depth $C$. 
In Fig.~\ref{fig:dependency_on_potential_depth2}
we show zeros of Jost functions $\mathcal{J}_{\kappa}(E)$ in the spherical square well potentials
with different potential depth $C=-70, -60, \cdots, 0$ MeV  
for $\tilde{l}=2$: $p_{3/2}$ ($\kappa=-2$) and $f_{5/2}$ ($\kappa=3$) 
and $\tilde{l}=3$: $d_{5/2}$ ($\kappa=-3$) and $g_{7/2}$ ($\kappa=4$).
Results with pseudospin $\tilde{s}=\pm{1}/{2}$ are denoted as solid and open 
dots, respectively.
For simplicity the results are shown in the complex energy plane. 
In the PSS limit, that is, $C=V+S=0$, the zeros are paired off and 
each pair of states coincide with each other, 
which is consistent with the formal analysis. 
When the potential depth increases, the zeros move gradually to the up left corner,
i.e., both the energy and the width become smaller.
Meanwhile the paired pseudospin partners separate from each other 
which means that the PSS is broken.
However, the distance between the paired points is not big and the PSS is conserved approximately. 

\begin{figure}
\begin{centering}
\includegraphics[width=0.9\columnwidth]{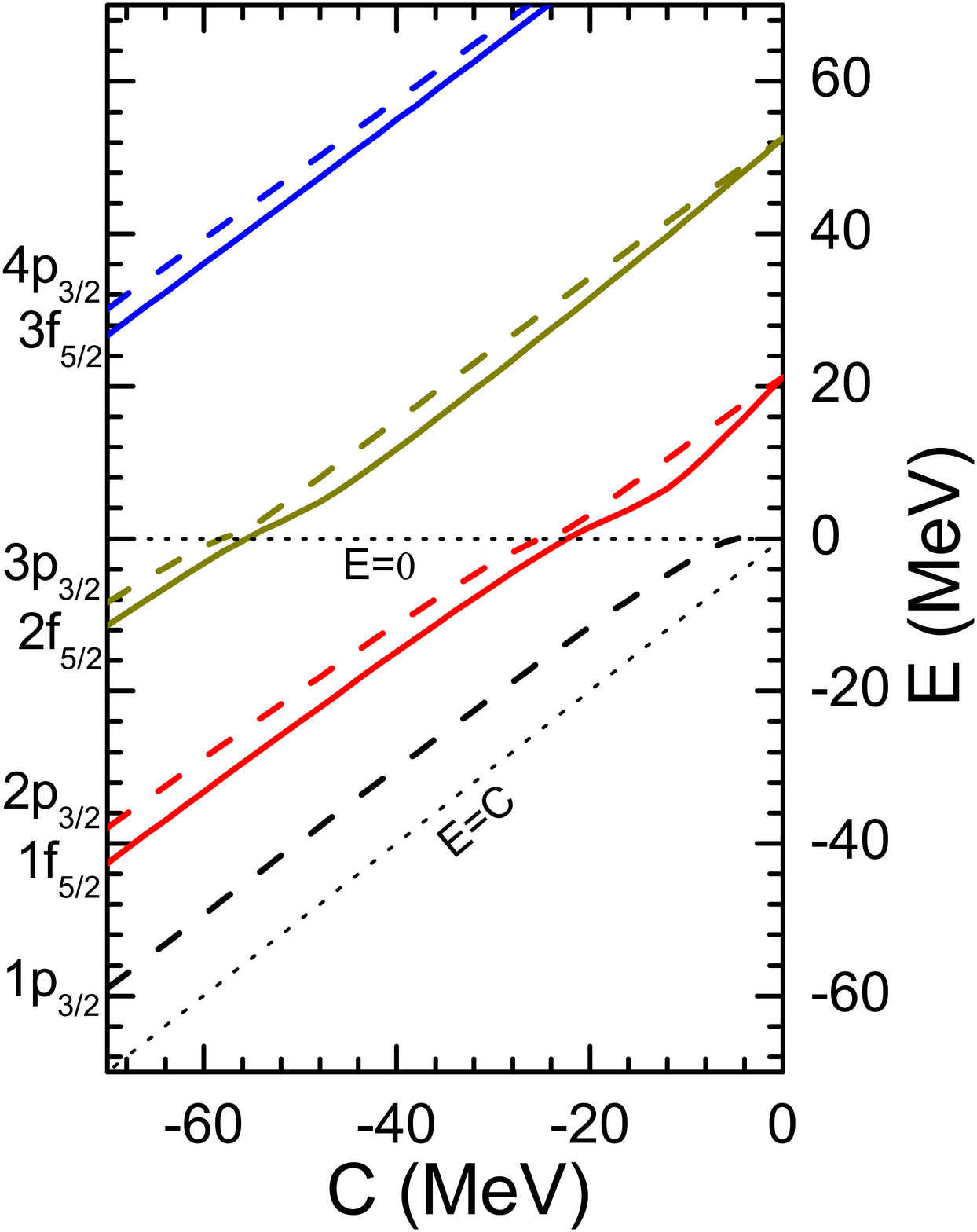}
\par\end{centering}
\caption{\label{fig:Efunc}(Color online).
Energies of bound and resonant states for $p_{3/2}$ and $f_{5/2}$ with $\tilde{l}=2$
in spherical well potentials as functions of the potential depth $C$. 
The results with pseudospin $\tilde{s}=\pm{1}/{2}$
are denoted as solid and dashed curves, respectively. 
All the levels are paired off except the lowest one. 
The bottom of the potential $E=C$ and the bound state threshold $E=0$ 
are shown as dotted lines.
}
\end{figure}

In Fig.~\ref{fig:Efunc} we show the calculated energies of bound and resonant
states with $\tilde{l}=2$ in the square well potentials as functions of the potential
depth $C$. 
As the potential depth varies from 0 MeV to $-$70 MeV, the energies are always paired
off except the lowest one. 
In the PSS limit, i.e., $C=0$, there is no bound state and
all the levels are resonant states with finite widths.
The PSS is exactly manifested in this case.
When the single particle potential becomes deeper,
the energies of both pseudospin partners decrease 
and the energy splitting first increases then decreases. 
After one of the pseudospin partners becomes a bound state, 
an interesting phenomena appears, that is,
one level is still a resonant state while another level becomes
to be bound. 
Due to the PSS, their energies are almost
the same, except that one is a little bit higher than the
threshold while another a little bit lower. 
Because a resonant state is very close to the threshold, 
the corresponding width is also rather small. 
In other words, the pseudospin partner of a single particle bound state 
may be another bound state or a ``quasi-bound'' state with very long life time.
When the depth of the potential increases further,
both of the PSS partners become bound states.
In this case we can discuss the usual PSS for bound states.

There have been some investigations concerning  
the dependence of the PSS in resonant states on parameters, e.g., the depth, the radius, 
and the diffuseness, of Woods-Saxon potentials~\cite{Guo2005_PRC72-054319,*Guo2006_PRC74-024320}.
Different from those studies, now by examining the Jost function,
we are able to trace the PSS partners from the case of finite potential
depth to the PSS limit continuously. 
From this point of view, 
it is more helpful in understanding the origin and splitting of the PSS.

\begin{figure}
\begin{centering}
\includegraphics[width=0.9\columnwidth]{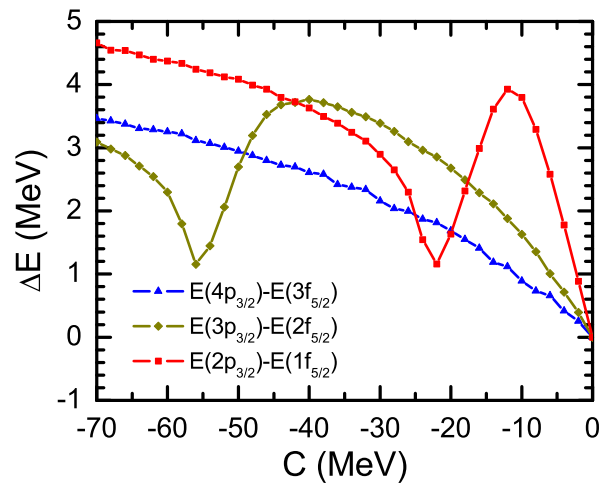}
\par\end{centering}
\caption{\label{fig:Delta_E}(Color online).
The energy splitting between PSS partners with the pseudo orbital 
angular momentum $\tilde{l}=2$, $p_{3/2}$ and $f_{5/2}$,  
as a function of the potential depth $C$.
}
\end{figure}

The energy splitting between PSS partners with the pseudo orbital 
angular momentum $\tilde{l}=2$, $p_{3/2}$ and $f_{5/2}$, is shown  
as a function of the potential depth $C$ in Fig.~\ref{fig:Delta_E}.
For simplicity only results for the lowest three pairs are shown here. 
First let us focus on the splitting of the lowest PSS pair, the levels 2$p_{3/2}$ and 1$f_{5/2}$. 
When the potential depth increases from 0, the energy splitting
first increases then decreases until they encounter the threshold at
a critical value of $C$ where one of the levels becomes a bound state
and the splitting takes a minimum value.
When the potential becomes even deeper, the splitting increases again.
This kind of threshold effect is also observed for other PSS pairs except that 
the critical value of $C$ is different.
The origin of this threshold effect should be one of the future topics
concerning the PSS in resonant states.


\begin{figure}
\begin{centering}
\includegraphics[width=0.9\columnwidth]{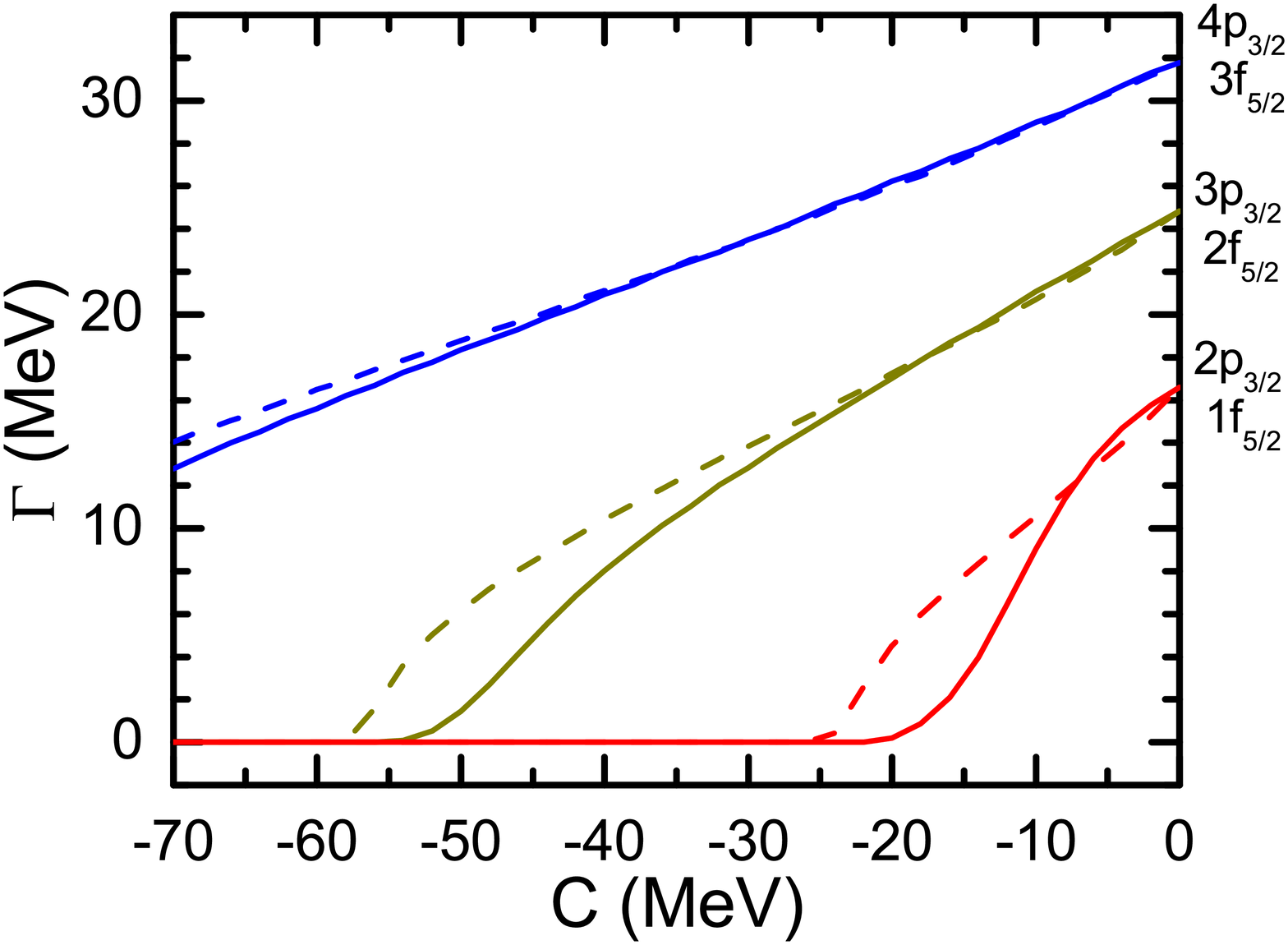}
\par\end{centering}
\caption{\label{fig:Gammafunc}(Color online).
Widths of resonant states for $p_{3/2}$ and $f_{5/2}$ with $\tilde{l}=2$
in spherical well potentials as functions of the potential depth $C$. 
The results with pseudospin $\tilde{s}=\pm{1}/{2}$
are denoted as solid and dashed curves, respectively. 
}
\end{figure}

\begin{figure}
\begin{centering}
\includegraphics[width=0.9\columnwidth]{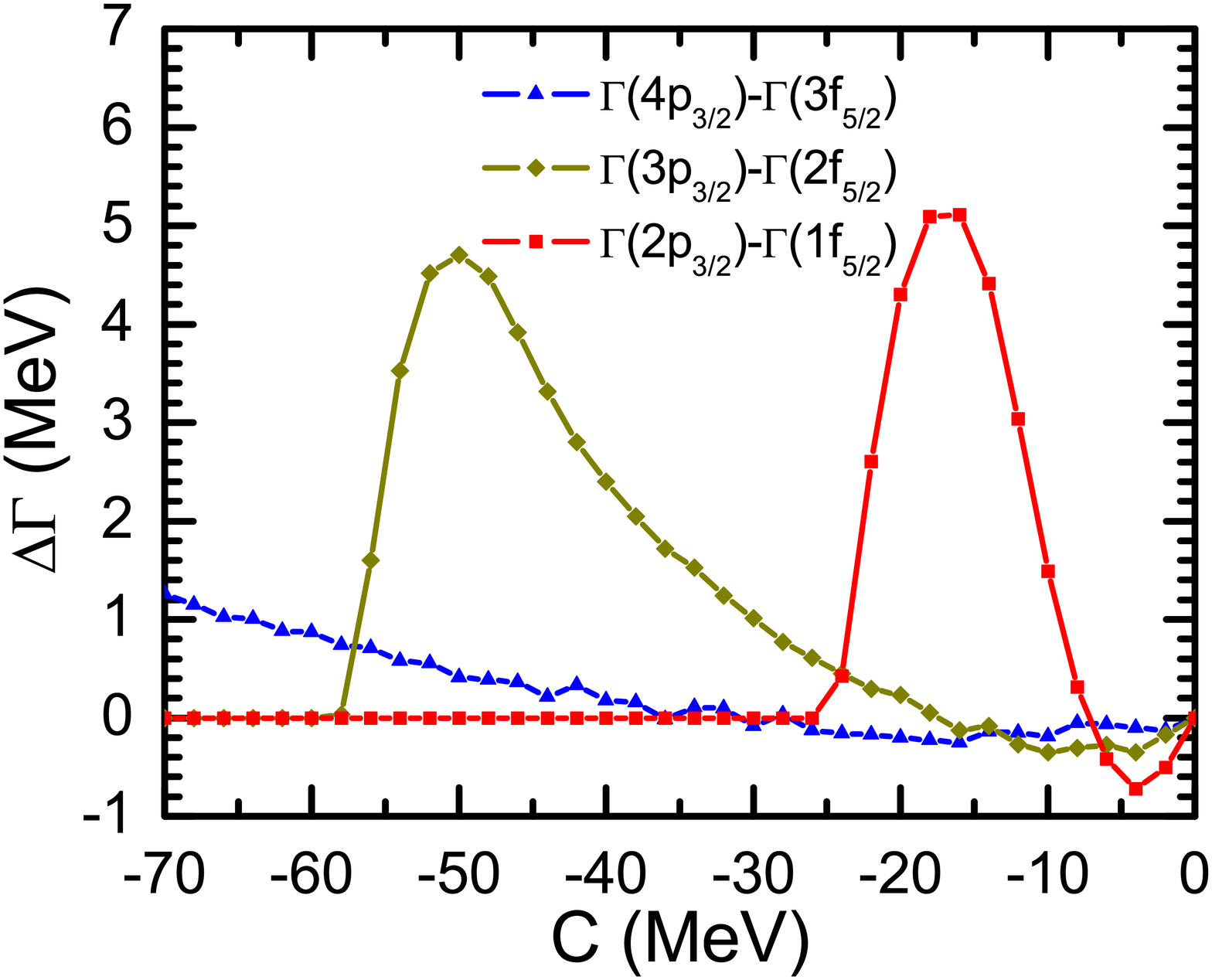}
\par\end{centering}
\caption{\label{fig:Delta_Gamma}(Color online) 
The width splitting between the PSS partners with the pseudo orbital angular momentum $\tilde{l}=2$,
$p_{3/2}$ and $f_{5/2}$, 
as a function of the potential depth $C$. 
}
\end{figure}

In Fig.~\ref{fig:Gammafunc} we show the calculated widths
of resonant states with $\tilde{l}=2$ in square well potentials 
as functions of the potential depth $C$. 
In this figure a zero width means the corresponding state has become a bound state.
Similar to the case of energies, the widths are also paired off when $C=0$.
Starting from the PSS limit, the width is a monotonically decreasing function of
the potential depth.
When the potential depth is finite, widths of a pair of pseudospin doublets are
different. 
First the level with $\tilde{s}=1/2$ is wider. 
But at a critical value of $C$, there occurs a crossing between 
the two curves at which the two widths are the same.
When the potential depth becomes even larger, 
the level with $\tilde{s}=-1/2$ becomes wider and the difference between two widths 
increases until it reaches a maximum value.
Afterwards, the width splitting decreases until it becomes zero
when both states become to be bound.  
This anomalous variation tendency can be seen more clearly in Fig.~\ref{fig:Delta_Gamma}
in which the width difference between PSS partners, 
$\Delta\Gamma \equiv \Gamma_{\tilde{s}=-1/2} - \Gamma_{\tilde{s}=+1/2}$,
is presented.
The width splitting first decreases from zero to a maximum value with
negative sign, then it increases and becomes zero. 
After the inversion of the width splitting, the splitting increases
and reaches a maximum value, then it becomes smaller again.
For each PSS pair, the width splitting assumes its largest value
at some point above the corresponding threshold.
%

Up to now, we have examined the dependence of the PSS 
on the depth of the single particle potential. 
From Eq.~(\ref{eq:Jost_function}) one can also learn the dependence of the
PSS on the radius of the potential $R$:
With $R$ increasing, the splitting term becomes smaller and the PSS
is more conserved.
There is no diffuseness in a square well potential. 
It would be more useful to get analytic results for more realistic potentials.
Then one can discuss the PSS in single particle resonant states in more realistic potentials.

\section{\label{sec:summary}Summary and perspectives}

The pseudospin symmetry in single particle resonant states is 
analyzed in details in spherical square well potentials. 
We built the Jost function for the small component of 
the Dirac wave function and studied 
resonant states 
as well as 
bound states 
by examining the zeros of the Jost function. 
The exact conservation and breaking of the PSS are investigated
and a novel way is used to show the origin of the appearance of intruder states:
The lowest zero of the Jost function with $\tilde{s}=-1/2$ 
is always isolated while the others are paired off with those of the Jost function
with $\tilde{s}=+1/2$.
When parameters of square well potentials are fixed, 
we studied the dependence of the PSS on the energy and $\tilde{l}$ of pseudospin doublet states.
It is found that the energy splitting is larger for higher pseudo
orbit angular momentum and at very high energies the splittings
between PSS partners are negligible. 
By examining the Jost function, we are able to trace continuously the PSS partners 
from the PSS limit to the case with a finite potential depth. 
As the depth of the single particle potential becomes deeper, 
the exact PSS begins to be broken and  
a threshold effect in the energy splitting is found:
The energy splitting first increases then decreases until the pseudospin doublets encounter 
the threshold where one of the levels becomes a bound state
and the splitting takes a minimum value; 
When the potential becomes even deeper, the splitting increases again.
When the depth of the single particle potential increases from zero,
there appears an anomaly in 
the width splitting of PS partners: It first decreases from zero to a maximum value with
negative sign, then increases and becomes zero again; 
after the inversion of the width splitting, the splitting increases
and reaches a maximum and positive value, then it becomes smaller and finally reaches zero.

Finally we would like to mention that the work presented
in Ref.~\cite{Lu2012_PRL109-072501} and here extends 
the study of relativistic symmetries to resonant states.
Although we have addressed several issues concerning the exact and
breaking of the PSS in single particle resonant states, there are still many 
open problems~\cite{Lu2013_AIPCP1533-63}, we list several of them below: 
\begin{itemize}
\item
Are there any experimental evidences of the PSS or SS in single particle resonant states?
\item
Having in mind that the centrifugal barriers are quite different
for pseudospin doublets of single particle resonant states, 
how to understand intuitively that their widths are exactly the same in the PSS limit?
\item
Is the threshold effect found in the energy splitting of PS partners in spherical
square well potentials general for other potentials? If yes, what is its origin?
\item 
Is the anomaly found in the width splitting of PS partners in spherical
square well potentials general for other potentials? If yes, what is its origin?
\item
What about the relations between Jost functions of pseudospin partners?
\item
How about the PSS or SS in resonant states in anti-nucleon or anti-hyperon spectra?
\item
How about the PSS or SS in resonant states in deformed systems?
\item
What is the effect of the Coulomb interaction (e.g., for protons) on the PSS
or SS symmetries?
\end{itemize}


\begin{acknowledgments}
Helpful discussions with R. V. Jolos, Haozhao Liang, Jie Meng, P. Ring, Jiang-Ming Yao, 
and Bin-Song Zou are acknowledged. 
This work has been supported by Major State Basic Research Development 
Program of China (Grant No. 2013CB834400), 
National Natural Science Foundation of China (Grants No. 11121403, No. 11175252, 
No. 11120101005, No. 11211120152, and No. 11275248), 
the Knowledge Innovation Project of Chinese Academy of Sciences (Grant No. KJCX2-EW-N01). 
The results described in this paper are obtained on the ScGrid of Supercomputing
Center, Computer Network Information Center of Chinese Academy of Sciences.
\end{acknowledgments}


%

\end{document}